
\magnification\magstephalf 
\baselineskip13pt
\vsize23.5truecm
\hoffset-0.75truecm
\voffset0.50truecm
\tolerance400

\newdimen\fullsize	
\fullsize=17.0truecm \hsize=8.0truecm 

\def\fullline{\hbox to \fullsize}
\let\lr=L \newbox\leftcolumn
\output={\if L\lr
		\global\setbox\leftcolumn=\columnbox \global\let\lr=R
	\else \doubleformat \global\let\lr=L\fi
	\ifnum\outputpenalty>-20000 \else\dosupereject\fi}
	\def\doubleformat{\shipout\vbox{\makeheadline
		\fullline{\box\leftcolumn\hfil\columnbox}
		\makefootline}
	\advancepageno}
	\def\columnbox{\leftline{\pagebody}}

\supereject
\if R\lr \null\vfill\eject \fi

\def\hskipcertainamount{\hskip8.10truecm}
\footline={\hskipcertainamount\rm -- \folio\ -- \hss}
\def\makefootline{\baselineskip24pt
	\line{\the\footline}}

\def\d{{\rm d}}
\def\E{{\rm E}}
\def\sumin{\sum_{i=1}^n}
\def\dell{\partial} 
\def\arr{\rightarrow}
\def\normal{{\cal N}}
\def\hatt{\widehat}
\def\tilda{\widetilde}
\def\half{\hbox{$1\over 2$}}

\font\csc=cmcsc10 
\font\smallsl=cmsl8 

{{
\hsize17.0truecm
\centerline{\bf  
	MINIMUM L2 AND ROBUST KULLBACK--LEIBLER ESTIMATION$^*$\footnote{}
	{\smallsl\hskip-20pt 
	$^*$From Proceedings of the 12th Prague 
 	Conference, 1994}}

\smallskip
\centerline{\bf Nils Lid Hjort, University of Oslo}
\centerline{\bf Department of Mathematics, 
	N--0316 Oslo, Norway} 
}}

\hsize8.0truecm 

\bigskip
{{\smallskip
\baselineskip12pt\sl 
{\bf Abstract.} 
This paper introduces two new robust methods 
for estimation of parameters in a given parametric family. 
The first method is that of `minimum weighted L2', 
effectively minimising an estimate of the integrated 
(and possibly weighted) squared error.
The second is `robust Kullback--Leibler',
consisting of minimising a robust version of the empirical 
Kullback--Leibler distance, and can be viewed as a general 
robust modification of the maximum likelihood procedure.  
This second method is also related to 
recent local likelihood ideas for semiparametric density estimation. 
The methods are described, influence functions are found,
as are formulae for asymptotic variances. In particular 
large-sample efficiencies are computed under the home turf conditions 
of the underlying parametric model.  
The methods and formulae are illustrated for the normal model.  
\smallskip}}

\bigskip
{\bf 1. Minimum weighted L2 estimation.}
Let $X_1,\ldots,X_n$ be independent data points from 
an unknown density $f$, and suppose that 
the data are to be fitted to some given 
regular parametric family of densities $f_\theta(x)$. 
A simple and natural estimation idea is to minimise an estimate of 
$\int w(f_\theta-f)^2\,\d x$, where $w(.)$ is a suitable weight function,
perhaps the constant 1. Disregarding the one term that does not 
depend on the parameter, this leads to the following strategy:
minimise 
$$Q_n(\theta)
=\int wf_\theta^2\,\d x-2{1\over n}\sumin w(x_i)f_\theta(x_i). \eqno(1.1)$$  
Taking the derivative this is also the same as solving 
$$V_n(\theta)=\int wf_\theta u_\theta(\d F_n-f_\theta\,\d x)=0, \eqno(1.2)$$
where $u_\theta(x)=\dell\log f_\theta(x)/\dell\theta$ 
is the score function of the model, and where 
$F_n$ is the empirical distribution of the data. 
\eject 

~ 

\vskip1.75truecm 
We derived (1.2) as a consequence of the natural (1.1),
but forming an estimator by solving this second equation 
can also be motivated separately. It forces a weighted integral
of the nonparametric $\d F_n(x)$ to be equal to the 
corresponding weighted integral of the parametric $f_\theta(x)\,\d x$. 
In spite of much work in the literature on 
various minimum distance strategies, 
the particular estimator (1.1)--(1.2) does not seem to have been 
studied earlier. It has also been proposed independently by 
M.C.~Jones (personal communication). 
A method recently considered in Brown and Hwang (1993) 
has intentions similar to that of (1.1), 
but is unnecessarily hampered with an intermediate histogram approximation.
This is a case of `don't smooth if you don't have to'. 

\bigskip
{\bf 2. Influence function.}
Let $\hatt\theta$ be the estimator and let 
$\theta_0$ minimise $\int w(f_\theta-f)^2\,\d x$.
There is typically a unique parameter achieving this, 
and we interpret this $\theta_0$ as 
the `least false' or `most appropriate' parameter value. 
As $n$ grows $\hatt\theta$ converges almost surely to $\theta_0$. 
Standard Taylor arguments show that 
$$\hatt\theta-\theta_0\doteq 
	-V_n^*(\theta_0)^{-1}V_n(\theta_0), \eqno(2.1)$$
where $V_n^*(\theta_0)$ is the matrix of derivatives of $V_n(\theta)$. 
Letting $u_\theta^*$ be the matrix of second order derivatives of 
the log density we have 
$$\eqalign{V_n^*(\theta)
&=\int w(f_\theta^2u_\theta u_\theta'+f_\theta u_\theta^*)
	(\d F_n-f_\theta\,\d x) \cr
&\quad -\int wf_\theta^2u_\theta u_\theta'\,\d x, \cr}$$
so that 
$-V_n^*(\theta_0)\arr_p J=J(\theta_0)$, 
where 
$$\eqalign{J(\theta)
&=\int wf_\theta^2u_\theta u_\theta'\,\d x \cr
&\quad -\int w(f_\theta^2u_\theta u_\theta'+f_\theta u_\theta^*)
	(f-f_\theta)\,\d x. \cr}$$
The influence function of the estimator can now be 
established, via (2.1); see for example Huber (1981) 
for definition of and important uses of influence functions. 
Here it becomes 
$$I(f,x)=J^{-1}\{w(x)f(x,\theta_0)u(x,\theta_0)-\xi_0\}, \eqno(2.2)$$
where 
$$\eqalign{
\xi_0&=\E_fw(X)f(X,\theta_0)u(X,\theta_0) \cr
&=\int w(x)f(x,\theta_0)f(x)u(x,\theta_0)\,\d x \cr
&=\int w(x)f(x,\theta_0)^2u(x,\theta_0)\,\d x. \cr}$$
Where notationally convenient we write 
$f(x,\theta)$ for $f_\theta(x)$, and so on.  
The (2.2) function is typically bounded, which means robustness. 
The influence function is in fact also 
redescending in most cases, going to zero 
for $x$-values outside mainstream.
This is often considered an attractive robustness feature 
of an estimation method. 


\bigskip
{\bf 3. Limit distribution.}
By the central limit theorem and the definition of $\theta_0$,
$\sqrt{n}V_n(\theta_0)$ tends to $\normal\{0,M\}$, where 
$$\eqalign{M
&={\rm VAR}_f\{w(X)f(X,\theta_0)u(X,\theta_0)\} \cr
&=\int w^2f_\theta^2fu_\theta u_\theta'\,\d x-\xi_0\xi_0'. \cr}$$
\noindent From (2.1) follows 
$$\sqrt{n}(\hatt\theta-\theta_0)
	\arr_d\normal\{0,J^{-1}MJ^{-1}\}, \eqno(3.1)$$
with $J$ as given above. Note that this result has been reached 
without having to assume that the true $f$ belongs to the parametric model. 

The expressions for $J$ and $M$ 
simplify under model conditions. 
Of course there is some loss of efficiency, 
that is, the limiting covariance matrix $J^{-1}MJ^{-1}$ 
is larger than the best possible one under the model, 
namely $(\int f_\theta u_\theta u_\theta'\,\d x)^{-1}$, 
achieved by the maximum likelihood method. 

\bigskip
{\bf 4. Local and weighted L2 fitting.} 
The size of the limiting variances depend on the weight function
$w(.)$. Choosing a local weight function can be contemplated, 
say of the kernel type $K_h(x_0-t)$ around a given $x_0$.
Here $K_h(u)=h^{-1}K(h^{-1}u)$ and $K$ is a given kernel function.
This gives a locally estimated normal, for example, 
in a spirit similar to local likelihood methods discussed
in Hjort and Jones (1994).  
The apparatus above can be used to investigate 
influence functions and large-sample properties. 

It is sometimes desirable to let the weight function 
be data driven too, perhaps to increase precision 
under close to the model circumstan\-ces.  
One example would be to use 
$w_n(x)=w_0((x-\tilda\mu)/\tilda\sigma)$,
for a suitable $w_0(.)$ function, 
with preliminary robust estimates of location and scale.
Result (3.1) is still true under appropriate conditions,
with $J$ and $M$ being defined in terms of 
the limit function version of $w_n(.)$.  

\bigskip
{\bf 5. Local and robust Kullback--Leibler fitting.}
The local kernel smoothed likelihood function, around a given $x_0$, is
$$\eqalign{
L_n(x_0,\theta)
&=\sumin K_h(x_i-x_0)\log f(x_i,\theta) \cr
&\quad\quad
-n\int K_h(t-x_0)f(t,\theta)\,\d t, \cr} \eqno(5.1)$$
see Hjort and Jones (1994). 
As shown in Hjort and Jones (1994), maximising (5.1) aims 
at minimising the localised Kullback--Leibler distance 
$$\eqalign{
d(f,f_\theta)
&=\int K_h(t-x_0)\bigl[f(t)\log{f(t)\over f_\theta(t)} \cr
&\qquad\qquad\qquad	
	-\{f(t)-f_\theta(t)\}\bigr]\,\d t \cr}$$
\noindent from true density to parametric density. 
In other words, the maximiser of (5.1) 
aims at a `least false' parameter value $\theta_0$ 
that in general is different from 
the one associated with the minimum weighted L2 method.  
Note that a large $h$ gives a flat $K_h(t-x_0)$ function, 
and brings back the ordinary Kullback--Leibler distance and 
the traditional full likelihood method.

The aim of Hjort and Jones (1994) is primarily the complete 
semiparametric estimation of the full density curve, 
as partly opposed to concentrating on the locally estimated 
parameters themselves. But this is also automatically one way
of obtaining robust parameter estimates for a given 
parametric family: Apply the above for a suitable 
centrally placed $x_0$, for a reasonably sized $h$.
The resulting maximiser $\hatt\theta$ is a robust estimate of $\theta$,
and $f(x,\hatt\theta)$ a robust estimate of the underlying density curve. 

Hjort and Jones (1994) demonstrate that 
$$\sqrt{n}(\hatt\theta-\theta_0) 
	\arr_d\normal\{0,J_h^{-1}M_hJ_h^{-1}\}, $$
with certain generally valid expressions 
available there for $J_h$ and $M_h$. 
At the moment it will suffice to give these under model conditions:
$$\eqalign{
J_h&=\int K_h(t-x_0)u_\theta u_\theta' f_\theta\,\d t, \cr
M_h&=\int K_h(t-x_0)^2u_\theta u_\theta' f_\theta\,\d t
		-\xi_0\xi_0', \cr}\eqno(5.2)$$
where $\xi_0=\int K_h(t-x_0)u_\theta f_\theta\,\d t$. 
The influence function of this robustified maximum likelihood 
is also derived in Hjort and Jones (1994), and is of the form
$$I(f,x)=J_h^{-1}\{K_h(x-x_0)u(x,\theta)-\xi_0\}. \eqno(5.3)$$
This is reasonably similar to the influence function (2.2) 
for the minimum L2 method. In many cases the present method,
with a suitably chosen $h$, is more efficient at the model
than at least the unweighted version of the minimum L2 method.  

\bigskip
{\bf 6. The normal model.}
The most important special case is that 
of fitting data to a normal $(\mu,\sigma^2)$. 

\medskip
{\csc 6.1. The minimum L2 method.} From (1.2) two equations 
are easily put up to 
define minimum L2 estimators $\hatt\mu$ and $\hatt\sigma$. 
These are solved for example by the iterative 
Newton--Raphson technique. 
Regarding performance, under Gau\ss ian circumstances, 
and using a constant weight function, we find  
$$\eqalign{
J&=(\sigma^3\sqrt{2\pi})^{-1}{\rm diag}(1/2^{3/2},3/2^{5/2}), \cr
M&=(\sigma^4 2\pi)^{-1}{\rm diag}(1/3^{3/2},2/3^{3/2}-1/8). \cr}$$
This gives an asymptotic variance for $\hatt\mu$ 
of size $1.5396\,\sigma^2/n$ and an asymptotic variance for $\hatt\sigma$ 
of size $0.9241\,\sigma^2/n$. These should be compared to 
the minimum possible values, under model conditions. 
These optimal figures are achieved by the ML method, 
and are $\sigma^2/n$ and $\half\sigma^2/n$, respectively. 
This makes the direct minimum L2 method qualify as 
a `quite robust but perhaps too inefficient method'. 
Increased efficiency at the model is achieved 
through appropriate choices of weight function $w(.)$,
cf.~comments at the end of Section 4. 
One possibility here is 
$w_n(x)=\exp\{\half\delta(x-\tilda\mu)^2/\tilda\sigma^2\}$,
defined in terms of preliminary robust estimates of location and scale,
and with an extra tuning parameter $\delta\in(0,1)$. 
Choosing e.g.~$\delta=0.8$ leads to quite good efficiency
at the model, while still retaining a reasonable robustness. 

\medskip
{\csc 6.2. The robustified ML method.} 
The robust Kullback--Leibler fitting method of Section 5 
can easily be made better than the unweighted minimum L2 method.  
For the present normal model, let us use a normal kernel.  
The method is then to minimise, for given $x_0$, the function 
$$\eqalign{
{1\over n}&\sumin \phi\Bigl({x_i-x_0\over h}\Bigr){1\over h}
	\{\log\sigma+\half(x_i-\mu)^2/\sigma^2\} \cr
&+\phi\Bigl({x_0-\mu\over \sqrt{\sigma^2+h^2}}\Bigr)
	{1\over \sqrt{\sigma^2+h^2}} \cr} \eqno(6.1)$$
over all $(\mu,\sigma)$. We may compute 
the $J_h$ and $M_h$ matrices of (5.2) 
without serious difficulties. But in the present context 
the interest lies more in getting hold of a single, 
robust $(\mu,\sigma)$-estimate,
than in obtaining a full function of local estimates. 
Therefore we suggest using $x_0=\tilda\mu$, a robust preliminary 
estimate of the mean, say the simple median. Minimising 
(6.1) with this $x_0$ defines the proposed $\hatt\mu$ and $\hatt\sigma$. 

Again it is of interest to see how well the method fares under 
Gau\ss ian home-turf conditions. 
Somewhat arduous calculations give two diagonal matrices 
$(J_\mu,J_\sigma)$ and $(M_{\mu},M_{\sigma})$ for $J_h$ and $M_h$ 
of (5.2). Here 
$$J_\mu={1\over \sigma^2}{1\over \sqrt{2\pi}}{1\over h}{1\over R^3}
\quad {\rm and} \quad 
M_\mu={1\over \sigma^2}{1\over 2\pi}{1\over h^2}{1\over S^3}, $$
in which 
$$R=(1+\sigma^2/h^2)^{1/2}
  \quad {\rm and} \quad 
  S=(1+2\sigma^2/h^2)^{1/2}. $$
Similarly, 
$$\eqalign{
J_\sigma&={1\over \sigma^2}{1\over \sqrt{2\pi}}{1\over h}
	\Bigl(1-{2\over R^2}+{3\over R^4}\Bigr), \cr
M_\sigma&={1\over \sigma^2}{1\over 2\pi}{1\over h^2}
\Bigl\{{1\over S}\Bigl(1-{2\over S^2}+{3\over S^4}\Bigr) \cr
&\qquad\qquad
	-{1\over R^2}\Bigl(1-{1\over R^2}\Bigr)^2\Bigr\}. \cr}$$
Thus $\sqrt{n}(\hatt\mu-\mu)\arr_d\normal\{0,\kappa_\mu^2\}$
and $\sqrt{n}(\hatt\sigma-\sigma)\arr_d\normal\{0,\kappa_\sigma^2\}$,
where the asymptotic variances are found as 
$M_\mu/J_\mu^2$ and $M_\sigma/J_\sigma^2$. 
Some calculations give 
$$\kappa_\mu^2
=\sigma^2{R^6\over S^3}
=\sigma^2{(1+1/k^2)^3\over (1+2/k^2)^{3/2}}, $$
writing $h=k\sigma$, and similarly 
$$\eqalign{\kappa_\sigma^2
&=\sigma^2{(1+1/k^2)^2\over (1+2/k^2)^{5/2}} \cr
&{(1+1/k^2)^3 (2+4/k^4)-(1+2/k^2)^{5/2}1/k^4
	\over (2+1/k^4)^2}. \cr}$$

How large should $h$ be chosen? We think of $h$ as $k\tilda\sigma$,
where $\tilda\sigma$ is a robust preliminary estimate of 
standard deviation, and need to choose the factor $k$. 
As a mild surprise the value $k=1$ gives precisely the 
same large-sample variances under the model as the straightforward
minimum L2 method of Section 2, respectively $1.5396\,\sigma^2$
and $0.9241\,\sigma^2$. A more efficient but still quite robust
value would be $k=2$, `place a normal with two standard deviations
around the median and maximise the local kernel smoothed likelihood'. 
Then the values are $1.063\,\sigma^2$ and $0.563\,\sigma^2$,
only a few percent above the values that are optimal under the
model, viz.~$\sigma^2$ and $\sigma^2/2$. 
Increasing the value to three estimated standard deviations 
brings the large-sample variances further down to
$1.015\,\,\sigma^2$ and $0.5152\,\sigma^2$.
One should not go much further if robustness is aimed for,
but of course a large $h$ gives back these optimal values. 

Comparing the performance of the weighted L2 method,
say with the data driven weight function indicated above, 
with that of the robust Kullback--Leibler estimator, 
is an interesting problem for further research. 
One should also devise criteria for choosing the necessary 
fine-tuning parameters. 

\bigskip
{\bf 7. Robust estimation of location and covariance matrix.}
The ideas and results above generalise easily to the multi-dimensional case.
In particular the localised Kullback--Leib\-ler method 
seems to constitute a fruitful way of obtaining
robust estimates of $\mu$ and $\Sigma$, 
the mean vector and covariance matrix of the underlying distribution. 
The estimates can be viewed
as robust estimates of these parameters under normality assumptions
but also outside normality. 

One concrete version of this scheme, 
in the $p$-dimensional case, is as follows:
Start out with preliminary and robust estimates 
$\tilda\mu$ and $\tilda\Sigma$ for mean and covariance matrix.
Then carry out local likelihood estimation with 
a Gau\ss ian kernel function centred at $\tilda\mu$ and 
with covariance matrix of size $h^2\tilda\Sigma$. 
This is seen to be the same as minimising the criterion function 
$$\eqalign{
\Bigl[{1\over n}&\sumin 
{\exp\{-\half(x_i-\tilda\mu)'\tilda\Sigma^{-1}(x_i-\tilda\mu)/h^2\} 
	\over h^{p}|\tilda\Sigma|^{1/2}} \cr
&\{\half\log|\Sigma|+\half(x_i-\mu)'\Sigma^{-1}(x_i-\mu)\}\Bigr] \cr
&+{\exp\{-\half(\mu-\tilda\mu)'
	(h^2\tilda\Sigma+\Sigma)^{-1}(\mu-\tilda\mu)\} 
\over |h^2\tilda\Sigma+\Sigma|^{1/2}} \cr} $$
over all possible $(\mu,\Sigma)$. 
Note that this method properly generalises that of (6.1). 
For $h$ larger than say~5 the procedure is 
practically the same as ordinary maximum likelihood estimation. 
A value of perhaps $h=2$ constitutes 
a modified maximum likelihood procedure 
with quite good robustness qualities without sacrificing 
much in efficiency under multinormal conditions. 

\bigskip
{\bf Acknowledgements.} 
This paper has benefited from ongoing joint work on related matters 
with M.C.~Jones, I.R.~Harris and A.~Basu. 

\bigskip
\centerline{\bf References} 

\parindent0pt
\baselineskip11pt
\medskip 

\def\ref#1{{\noindent\hangafter=1\hangindent=20pt
  #1\smallskip}}          

\ref{%
Brown, L.D.~and Hwang, J.T.G. (1993).
How to approximate a histogram by a normal density. 
{\sl American Statistician} {\bf 47}, 251--255.}

\ref{%
Hjort, N.L.~and Jones, M.C. (1994).
Locally parametric nonparametric density estimation.
Submitted for publication.}

\ref{%
Huber, P.J. (1981).
{\sl Robust Statistics.}
Wiley, New York.}

\eject 

\bye